# Single electron tunneling as a possible conduction mechanism in diamond-like carbon film

S. S. Tinchev[a], S. Alexandrova[c], E. Valcheva[b]

[a] Institute of Electronics, Bulgarian Academy of Sciences, Tzarigradsko Chaussee 72, 1784 Sofia, Bulgaria

[b] Physics Department, Sofia University, J. Bourchier 5, 1164 Sofia, Bulgaria

[c] Institute of Solid State Physics, Bulgarian Academy of Sciences, Tzarigradsko Chaussee 72, 1784 Sofia, Bulgaria

Abstract:

Nonlinear current–voltage characteristics and in some cases also step-like characteristics were observed in diamond like carbon films. We suggest that the transport mechanism is tunneling between conducting nanoparticles in the films and the observation of step-like structures is a manifestation of Coulomb blockade found for the first time in these films. We point out that the origin of similar current-voltage characteristics observed in other carbon structures can be also Coulomb blockade phenomenon.

Single electron tunneling phenomena such as Coulomb blockade have been attracting great interest because of their scientific and technological importance [1, 2]. Coulomb blockade of tunneling is a phenomenon observed in low capacitance tunnel junctions usually at low temperatures. Tunneling of a single electron causes charging of the junction capacitance and thus blocking the tunneling for voltages below $e/C$. Here $e$ is the electron charge and $C$ is the junction capacitance. In order to observe Coulomb blockade the charging energy $E_c=e^2/2C$ should be as large as the thermal energy $k_BT$ ($k_B$ is the Boltzmann constant and $T$ is the temperature). Therefore for structures with micrometer and even sub micrometer dimensions this phenomenon is observed typically at low temperatures below 1 K.

To fabricate single electron devices operating at room temperatures it is essential to obtain nano-sized quantum dots or nanostructures with sizes below 10 nm, because the charging energy can overcome the thermal energy ($k_BT$ = 26 mV at 300 K) only in such small structures. Dimensions below 10 nm are, however, below the resolution limit of the electron beam lithography. Therefore self-organization processes are promising candidates for preparing electron devices operating at room temperatures. In systems with metalic nanoclusters Coulomb blockade has already been observed at room temperature, see for example [3-8].

In this letter we discuss single electron tunneling as a possible conduction mechanism in diamond-like carbon (DLC) films at room temperature. As far as we know, this is the first time that Coulomb blockade is found in DLC films. The motivation for this work is the following. Diamond-like carbon films can be considered to consist of $sp^2$ - bonded conducting graphitic islands embedded in a $sp^3$ - bonded insulating matrix [9]. Estimating the capacitance between two graphitic clusters as spheres with equal diameter of about 3 nm at distance about 2 nm [10] one can find a value of about 0.5 aF. Thus in a tunnel junction with such small capacitance Coulomb blockade should be observed up to 3000 K. Therefore the first condition for room temperature observation of Coulomb blockade in diamond-like carbon is satisfied. There is, however, a second condition for observing single electron tunneling. The charging energy must be larger then the Heisenberg's uncertainty of the energy. From this it follows that the resistance of the junction R should be much larger then $\hbar/e^2 \approx 4.1$ k$\Omega$ ($\hbar$ being the Planck constant), which is also usually well fulfilled in high-resistance diamond-like carbon film.

Diamond-like carbon films used in our experiments were amorphous hydrogenated carbon films (a-C:H) made by DC PECVD from a mixture of benzene and argon. The films were deposited at a DC voltage of 1 kV and a pressure of $5.10^{-2}$ Torr. The deposition was carried out for 10 min and the thickness of the films was about 160 nm. The films were deposited on magnetron sputtered Ti films to serve as a bottom electrode. Top Ti-electrodes were deposited through a shadow mask with a circular aperture of a diameter of 1mm. Before every deposition 10 min Ar ion sputtering at 1 kV was carried out. The structures were not annealed.

The measurements of the I-V characteristics were carried out at room temperature in the dark. A Hewlett-Packard Model 4140 pA-meter/dc voltage source was used. Fig. 1 shows the current-voltage (I-V) characteristic of a Ti/(a-C:H)/Ti structure. The films display a nonohmic behavior. The I-V characteristic is nearly symmetrical for the negative and positive applied voltages. For small voltages this characteristic is almost linear (see the inset of Fig. 1), while for high voltages it is highly nonlinear and can be fit well to the Fowler - Northeim plot – Fig. 2. Our reasonable explanation of this behavior is tunneling – direct tunneling at small voltages, which is going to Fowler – Northeim tunneling with increasing voltage. Here one can point out that we excluded hopping as possible conduction mechanism, because the hopping conduction has linear I(V) behavior for all voltages in contrast to our observation of linear dependence only for small voltages.

Using the Simmons equation [11] for small voltages one can calculate the expected current density around zero voltage. The measured value, however, was many orders of magnitude smaller then the expected one. Similar small current density was measured also by other authors in DLC films and it is explained usually by a space charge limited current (SCLC) mechanism.

However, these properties can be understood in the framework of the model of collective charge transport in disordered arrays [12]. For one dimensional array of small conducting islands weakly coupled by tunneling this model predicts that the current is zero below a certain threshold voltage, which is proportional to the numbers of particles in the chain. Another prediction of [12] is that above the threshold voltage, the current follows a power low I~ $[(V-V_T/V_T)]^\zeta$, where $\zeta = 1$ for one-dimensional array and $\zeta \approx 2$ for two-dimensional array. Such behavior was observed indeed in artificial 1-D and 2-D arrays including in self-assembled chains of graphitized carbon nanoparticles [13] and in amorphous carbon dots array [14]. However, because of the particles with large dimensions used (diameter of 30 – 40 nm) single – electron effects, such as Coulomb blockade and Coulomb staircase were observed only at low temperatures ( 77 K and 9.4 K respectively).

Fig. 3 shows a part of the curve current vs. normalized voltage $(V-V_T)/V_T$ for our experimental data. Here the threshold voltage was set to 0.1 V. The observed behavior is almost linear, as it should be in 1-D array.

Based on our experiment a model can be suggested for explanation of the small electrical conductivity of amorphous hydrogenated carbon as shown in Fig. 4a. In this model of the DLC film tunneling is happens between nanometer size graphitic clusters inside the film. The film contains many conduction channels in parallel, which allows the passage of electrons from the bottom to the surface contact. Each channel is represented by a one-dimensional array of tunnel junctions. The corresponding equivalent circuit is shown in Fig. 4b, where every junction consists of a resistor $R$ in parallel with a capacitor $C$.

From the estimated capacitance of an isolated cluster of about 0.5 aF one can expect observation of Coulomb blockade for voltages below $U = e/C \sim 0.3$ V in good agreement with our measurements. As the number of junctions in the conduction channels increases when the film thickness increases, the Coulomb blockade regions also increase, so that the threshold voltage should be proportional to the film thickness. This is true, however, only for thin films with small number of junctions. For thicker films with many tunnel junctions between the top and bottom electrode, the conduction paths will form a three dimensional network and the threshold voltage will be no more proportional to the layer thickness. This is probably the situation in our case, where for thickness of 160 nm one can estimate about 30 serial connected junctions. In order to check it we measured also the current – voltage characteristics between two top contacts on the DLC surface (Fig. 5). The threshold voltage was almost the same although in this case the number of tunnel junctions connected in series doubled. This is evidence that this simple model can not describe adequately the experiment.

It should be noted that the channels are probably not entirely identical. There are an inevitable distribution of the cluster size as well as cluster distances and thus a dispersion of the individual threshold voltages of every channel. This can explain the rounding of the I-V characteristics at the onset of the current increase around the threshold voltage.

We would like to point out that in some cases step-like structures, known as Coulomb staircase, were observed in the I-V characteristic (see Fig. 5, curve 2). For existence of Coulomb staircase an asymmetry of the junction time constants RC is necessary. Obviously in these measurements this is the case. The most peaks in Fig. 5 are nearly equally spaced by about 0.9 V. From this value the cluster capacitance of $1.8 \times 10^{-19}$ F can be calculated and using the formula for capacitance of an isolated conducting sphere $C = 4\pi\varepsilon\varepsilon_0 r$ one can find the cluster diameter as 0.5 nm. This diameter should be regarded as estimated low limit of the cluster dimensions. However, additionally peaks at higher voltages are also seen, which is an indication for random distribution of size and separation of particles in the film. The observation of step-like structures is additional support of our explanation of these results by Coulomb blockade phenomenon.

Although the conductance is strongly suppressed around zero bias voltage, it still remains finite – inset on Fig. 1. This suggests the existence of elastic co tunneling in these films because I ~V inside the Coulomb blockade [15]. From this simple equivalent circuit one can estimate the tunnel resistance of a single junction of about $4 \times 10^{16}$ Ω, obviously high enough to ensure a long lifetime $\tau = RC \sim 20$ ms of trapped electron on an island before it tunnels out of the island and onto another electrode. In this calculation we use $4 \times 10^{10}$ channels placed in 1 mm diameter top contact and 30 serial connected junctions. Further support for our explanation of the observed behavior by the Coulomb blockade will be searched in future experiments involving a gate electrode to modulate the charge density as usually in single tunneling experiments.

It should be pointed out that usually the electrical conductivity of diamond-like carbon is explained by some other models: activated conductivity of delocalized states, hopping in band-tail states and at the Fermi level, multiphoton tunneling processes, as well as combination of these [16]. The mostly discussed mechanism is that of the variable-range hopping. However, it is well known that the electrical conductivity of DLC films can vary by many orders of magnitude depending on deposition conditions and obviously no single mechanism can explain the electrical conductivity of DLC film in all cases. Our simple model suggested above is applicable for high resistively DLC films and it is not in contradiction with these models because most of them are based on tunneling. In most measurements reported, however, the electrical conductance was measured at low electric fields. Therefore the Coulomb blockade can not be noticed. Instead, mainly the temperature dependence of the electrical conductivity has been measured and compared with the predictions of the different theoretical models for electrical transport in amorphous materials already mentioned above in order to distinguish between them. Only in [17] electrical resistance of DLC films in fields up to several MV/cm have been measured and found that it changes only slightly up to a field of $3 \times 10^5$ V/cm, but decreases sharply at higher fields. This value ~ 30 V/µm is of the same order of magnitude as our measured threshold voltage.

Similar current-voltage characteristics have been observed in electron emitting carbon films where also a threshold voltage is needed in order to have significant electron emission [18]. At the present there is no complete explanation of the field emission phenomena in diamond-like carbon films. In [19] four different mechanisms based on different models for charge transport were discussed: the Fowler-Nordheim model, the Schottky emission model, the space–charge-limited model and the Poole Frenkel mechanism. All four models were giving good fits to the emission data and it is not clear why one model should be used in preference to any other. The experimentally measured typical threshold electric field about 10V/µm is of the same order of magnitude as in our measurement. To our opinion the observed threshold voltage in electron emission from carbon films can be explained with Coulomb blockade.

Nanometer sized conductive clusters in a-C:H films were already observed in the past [20] by comparing AFM and STM pictures of the a-C:H films surface. While the AFM picture shows very smooth surface, the STM shows rough structure with nanometers sized conducting round hills. Recently $sp^2$- bonded nanoclusters with dimensions 1-3 nm were observed also in tetrahedral amorphous carbon films [21].

In summary, we have observed nonlinear current-voltage characteristics of diamond-like carbon films, which can be explained by tunneling between conducting graphitic clusters in the film. Additional support of this conclusion was given by the observation of step-like structures, which we interpreted as Coulomb staircase. Similar current-voltage characteristics observed in electron emission from diamond like carbon films can be also explained by Coulomb blockade. This observation could be a strong support for the cluster model of diamond-like carbon.

Figure captions:

Fig. 1. Current-voltage characteristic of a diamond-like carbon films measured between top and bottom electrodes. Inset shows the enlarged region of the Coulomb blockade at low voltages.

Fig. 2. A Fowler-Northeim plot of the I-V characteristic of Fig. 1 for high voltages. The strain line represents the ideal Fowler-Northeim relation $\ln(J/V^2) \sim 1/V$.

Fig. 3. A part of the curve current vs. normalized voltage $(V-V_T)/V_T$ for our experimental data.

Fig. 4. A model of DLC film with many conduction channels in parallel (a). Each channel is one-dimensional array of tunnel junctions. The corresponding equivalent circuit is (b).

Fig. 5. Current-voltage characteristics of a diamond-like carbon films measured between two top electrodes for positive voltages. In the case (2) step-like structures marked by arrows (Coulomb staircase) were observed.

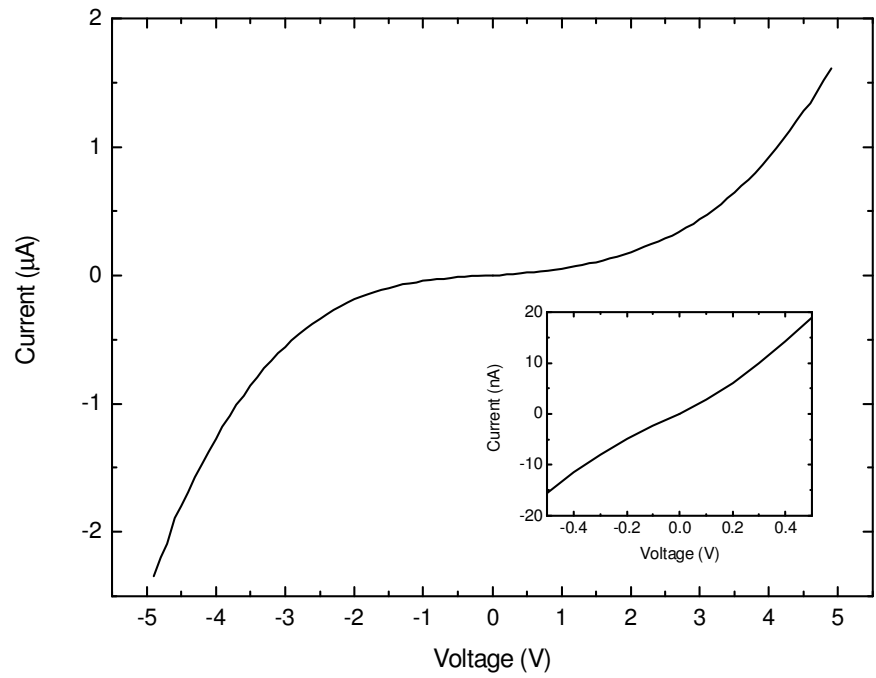

Fig. 1.

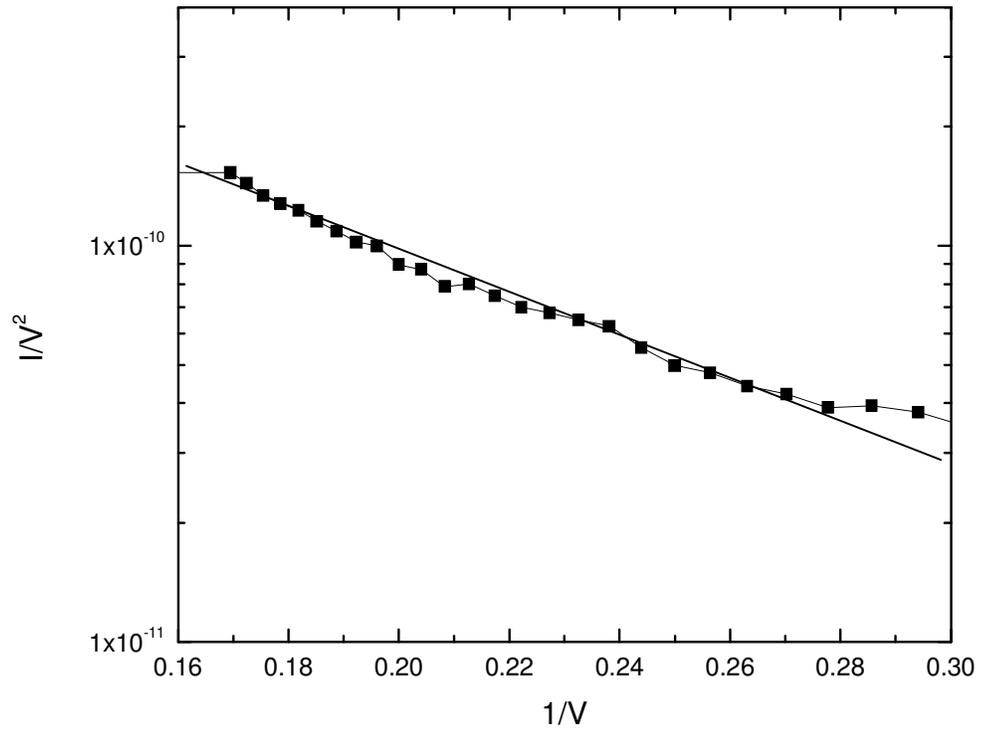

Fig. 2.

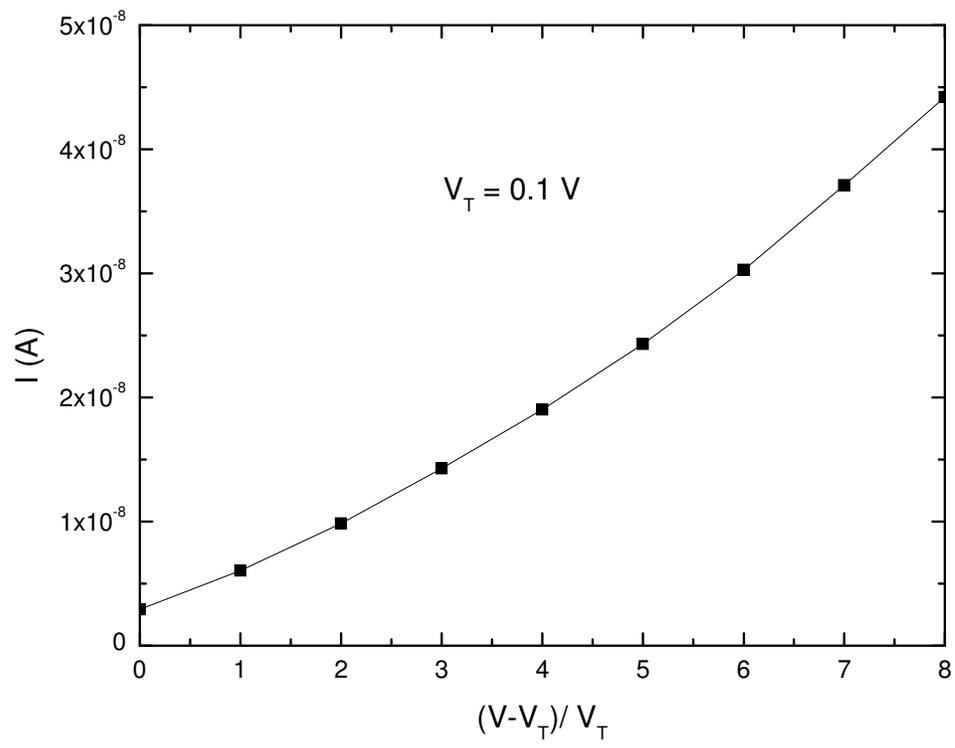

Fig. 3.

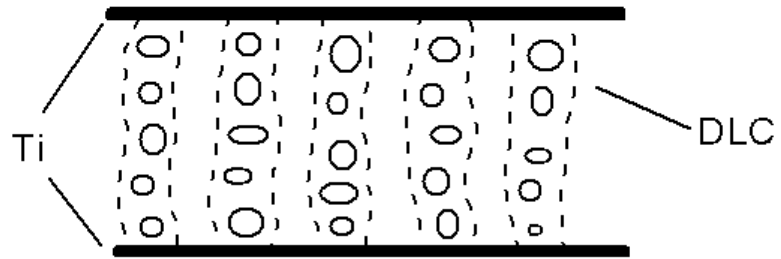

(a)

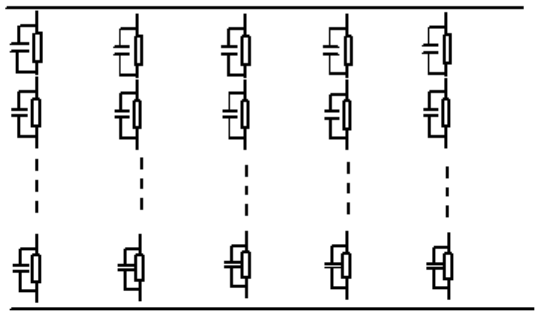

(b)

Fig. 4.

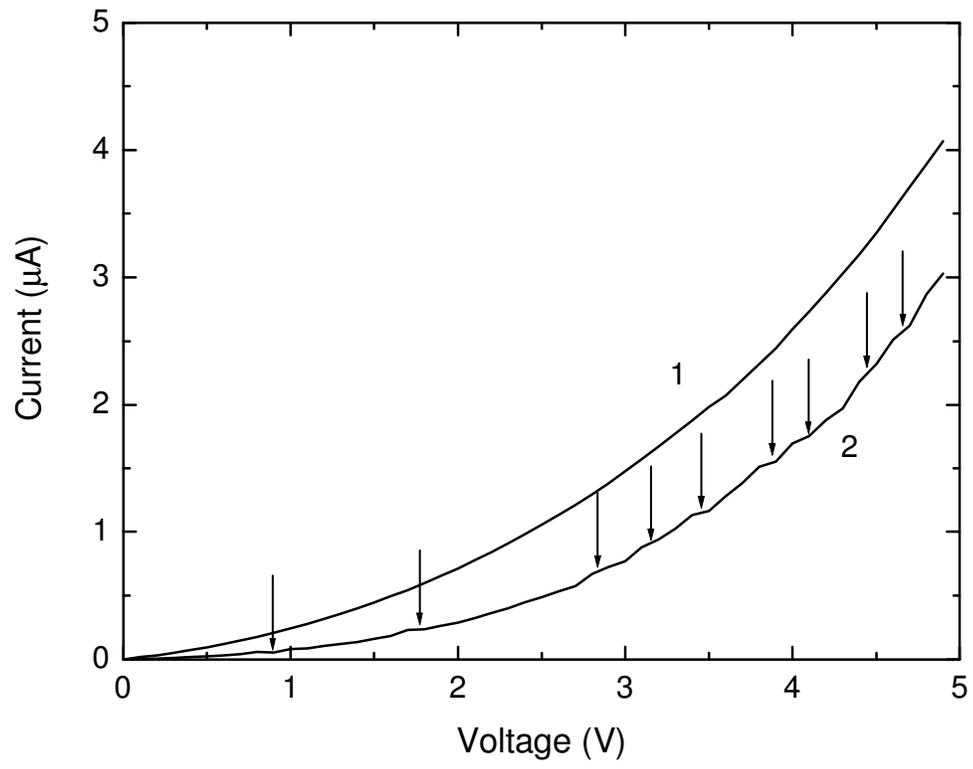

Fig. 5.